**A common rule for decision-making in animal collectives across species**

Sara Arganda[a,b,1], Alfonso Pérez-Escudero[a,1], Gonzalo G. de Polavieja[a,2]

[a]Instituto Cajal, Consejo Superior de Investigaciones Científicas, Av. Doctor Arce, 37, 28002 Madrid, Spain

[b]Present address: Centre de Recherches sur la Cognition Animale UMR 5169, Bât IVR3, 118 route de Narbonne F-31062 Toulouse cedex 09

[1]S.A. and A.P-E contributed equally to this work

[2]To whom correspondence should be addressed. E-mail: gonzalo.polavieja@cajal.csic.es

Gonzalo G. de Polavieja
Instituto Cajal, CSIC
Av. Doctor Arce, 37
Madrid 28002, SPAIN
Phone. +34 915854652
E-mail : gonzalo.polavieja@cajal.csic.es

**Short title:** General rule for collective decisions

**Abstract**

A diversity of decision-making systems has been observed in animal collectives. In some species, choices depend on the differences of the numbers of animals that have chosen each of the available options, while in other species on the relative differences (a behavior known as Weber´s law) or follow more complex rules. We here show that this diversity of decision systems corresponds to a single rule of decision-making in collectives. We first obtained a decision rule based on Bayesian estimation that uses the information provided by the behaviors of the other individuals to improve the estimation of the structure of the world. We then tested this rule in decision experiments using zebrafish (*Danio rerio*), and in existing rich datasets of argentine ants (*Linepithema humile*) and sticklebacks (*Gasterosteus aculeatus*), showing that a unified model across species can quantitatively explain the diversity of decision systems. Further, these results show that the different counting systems used by animals, including humans, can emerge from the common principle of using social information to make good decisions.

Sensory data always has some degree of ambiguity, so animals need to make decisions by estimating the properties of the environment from uncertain sensory data (1-5). This estimation has been shown to be close to optimal in many cases, making optimal Bayesian decision-making a successful framework shared by behavioral, neurobiological and psychological studies (1-7).

A richer scenario for decision-making takes place when animals move in groups. In this case, the behaviors of other animals are an extra source of information (6-34). Animals of different species have been observed to incorporate this extra information in their decisions in different ways. Some species make decisions that can be explained using the differences of the numbers of animals taking each option (21, 22), others according to the relative differences (Weber´s law) (23, 24) or using other rules (25-34). This diversity of decision schemes has translated into a diversity of models (21-22, 24-34).

To search for a unified framework having the diversity of decision-making schemes as particular cases, we generalized Bayesian decision-making to the case of animal collectives. Our previous attempt at building such a theory predicted that the only relevant social information is the difference of the numbers of individuals already choosing each available option, and not the numbers themselves or the relative differences (or Weber´s law) (22). But this theory was limited to the particular case in which only one of the options could be a good option (22). We have now generalized the theory, allowing all available options to be good or bad options. We found that this generalization explains the diversity of decision rules observed in collectives, maintaining the same conceptual and mathematical simplicity, and containing our previous theory as a particular case. We have tested the theory experimentally in decision experiments using zebrafish (*Danio rerio*), but to cover the diversity of decision systems, we have also tested it using rich datasets of decision-making in argentine ants (*Linepithema humile*) (24) and three-spined sticklebacks (*Gasterosteus aculeatus*) (25, 26). We found a quantitative match between the theory and the different decision systems of these representative species.

**Results**

We obtained how the behaviors of others should be taken into account to improve the estimations of the structure of the world and make decisions in animal collectives. For a situation with two identical options to choose from (**Fig. 1A**), we looked for the probability that one option, say $x$, is a good option given that $n_x$ and $n_y$ animals have already chosen options $x$ and $y$, respectively. We used Bayesian theory to find an approximated analytic expression for this probability as (see **Supporting Text**)

$$P(x \text{ is good}) = \frac{1}{1 + as^{-(n_x - k n_y)}} . \qquad (1)$$

Parameter $a$ measures the quality of non-social information available to the deciding individual, and $s$ measures how reliably an individual that has chosen $x$ indicates to the deciding individual that $x$ is a good option. According to **Eq. (1)**, the more individuals chose option $x$, $n_x$, the higher the probability that option $x$ is good for the deciding individual, and more so the higher the reliability $s$ of the information from the individuals that already chose $x$. On the other hand, each individual that chooses $y$ decreases the probability that $x$ is a good option. Parameter $k$ measures the relative impact of these two opposing effects. Individuals need to decide based on the estimated probabilities in **Eq. (1)**. A common decision rule in animals, from insects to humans, is probability matching, according to which the probability of choosing a behavior is proportional to the estimated probability (35-44),

$$P_x = \frac{P(x \text{ is good})}{P(x \text{ is good}) + P(y \text{ is good})}. \tag{2}$$

This rule is known to be optimal when there is competition for resources (39, 40) and when the estimated probabilities change in time (41-44). Probability matching in **Eq. (2)**, together with the estimation in **Eq. (1)**, gives that the probability of choosing $x$ is

$$P_x = \left(1 + \frac{1 + as^{-(n_x - k n_y)}}{1 + as^{-(n_y - k n_x)}}\right)^{-1}, \tag{3}$$

and $P_y = 1 - P_x$ is the probability of choosing $y$. The main implications of **Eq. (3)** are apparent in its plot, **Fig. 1B**. First, decision-making in collectives is predicted to be different for low and high numbers of individuals. For low numbers, there is a fast transition between preferring one side over the other, while for high numbers the transition has an intermediate region with no preference in which the probability has a plateau of value 1/2. There is a clear separation between the low and high numbers regimes at the point $\tau = \log(a)/(\log(s)(1-k))$ in which the plateau starts, (**Fig. 1B; see Supporting Text**). Second, in the high numbers regime, the isoprobability curves are straight lines of slope $k$. We can use this slope to classify three very different scenarios that we found correspond to different experimental data sets: $k = 0$, $0 < k < 1$ and $k = 1$, **Fig. 1C**.

For $k=0$, the animals at one option do not impact negatively on the estimated quality of the other option. This can take place, for example, when animals at one option do not seem to have information about the other option. An important prediction for this case is that for high number of animals there is a large plateau of probability 1/2 of choosing each of the two options (**Fig. 1C, left**). To have a significantly higher probability of choosing one option, say $x$, it is then needed, not only that $n_x > n_y$ but also to be outside of the large plateau, which means that very few animals have chosen the other option *y*, $n_y < \tau$. A second prediction is that there is a finite number of animals that need to be distinguished. To see this consider that the probability that option *x* is a good one, **Eq. (1),** for *k*=0 increases monotonically with $n_x$ and converges to 1. The number of animals $n_x$ needed to reach a high probability of 0.95 is given by $\alpha = (\log(a) - \log(1/0.95 - 1))/\log(s)$ (**Fig. S1**). Since beyond $\alpha$ the probability changes very little, in practice it is not necessary to count beyond that number. For a wide range of parameters *a* and *s*, $\alpha$ has low values, corresponding to counting up to a low number of animals (**Fig. S1**).

We have found that wildtype zebrafish, *Danio rerio*, in a two-choice set-up used for tests of sociability (45, 46) make choices that quantitatively correspond to the predictions of the *k*=0 case. The set-up has three chambers separated by transparent walls; a central chamber with the zebrafish we monitor, and two lateral chambers with different numbers of zebrafish acting as social stimuli (**Fig. 2A**; see **Materials and Methods**). An interesting feature of this set-up is that it measures the behavior of a single individual when presented with social stimuli, allowing a direct test of the individual decision rule in **Eq. (3)**. Specifically, we measured the probability that the focal fish chooses each of the two options for a range of configurations (**Fig. 2B**, each dot is the mean of typically *n*=15 animals). We found that these experimental results correspond to **Eq. (3)** for *a*=11.2, *s*=5 and *k*=0 (blue surface, **Fig. 2B**) with a robust fit (**Fig. S2**). To make a more quantitative comparison between theory and experiment, we highlighted several lines on the theoretical surface, using different colors to indicate different numbers of fish at option *y*. **Fig. 2C** compares the probability values for these five lines with the experimental data, showing a close match. The model offers both a quantitative fit to data and a simple explanation of the experimental result. Fish do not choose directly according to the number of other fish, but to how these numbers indicate that a place is a good option, giving a rule of 'counting up to three'.

The close match between experimental data and the decision-making model supports that zebrafish behavior corresponds to probabilistic estimations about the quality of sites using social information. However, the processing steps made by the fish brain need not have a one-to-one correspondence with the computational steps in the theory. Instead, a likely option is that zebrafish use simple behavioral rules that approximate good estimations. We found mechanistic models with simple probabilistic attraction rules for individual fish that approximate well the decision-making model and the data (**Figs. S3 and S4**).

The second case we consider has parameter $k$ in the range from 0 to 1. For this range, the estimation that $x$ is a good option increases with how many animals have already chosen $x$ and decreases, although at a slower rate, with how many have chosen option $y$. This situation might be common, for example, in food search. Animals choosing one option can indicate that there is a food source in that direction, but also that there might not be a food source at the other option. In this case, the probability of choosing $x$ has a plateau in which both options are equally likely, but increasing the number of animals that have chosen $x$, $n_x$, reaches a transition region of rapid increase in probability (**Fig. 1B**). This transition region follows a straight line of slope $k$ in the probability plot (**Fig. 1B**). This line obeys for high number of animals that $n_y \approx k\, n_x$. This is a Weber law (23, 24), according to which the just noticeable difference between two groups is proportional to the total number of individuals. Indeed, if we substitute $n_y \approx k\, n_x$ into $\Delta N / N \equiv (n_x - n_y)/(n_x + n_y)$ we obtain a constant of value $(k-1)/(k+1)$. A second prediction of the model is that decisions should deviate from Weber behavior at low numbers (below the transition point $\tau$ in **Fig. 1B**).

We have found that decisions made by the Argentine ant, *Linepithema humile*, correspond to the case $0 < k < 1$. Ants' choices to turn left and right have been recorded in reference (24) and we found that they have choice probabilities well described by **Eq. (3)**, except that experimental probabilities do not reach values as close to 0 or 1 as the theory. This difference might be due simply to the fact that ants are not always making turn decisions based on pheromone but responding to other factors like roughness of terrain or collisions with other ants. We therefore considered that ants choose at random with a given probability and otherwise make a decision according to **Eq. (3)** (see **Eq. (4)** in **Materials and Methods**). This modification only introduces an overall rescaling in the probabilities, so all structural features described below are present in **Eq. (3)** (see **Fig. S5**). We obtain a good correspondence with data for high (**Fig. 3A**) and low numbers of animals (**Fig. 3B**) with a fit that is robust (**Fig. S6**). The experimental data are smoother than the theory, without a central plateau, but still with a close correspondence, as also shown in the following analysis. According to Weber's law, isoprobability curves should be horizontal lines in the $\Delta N / N \equiv (n_x - n_y)/(n_x + n_y)$ versus $N \equiv n_x + n_y$ plane. This is true both for the theory and experiments for high numbers of total animals $N$, **Fig. 3C**. The advantage of this plot is that it magnifies the region of low $N$, where the data deviate from Weber's law similarly to the theoretical prediction. A further quantitative analysis revealing the close correspondence between theory and data is shown in **Fig. 3D**. We performed a linear fit to the experimental probability along the lines of constant $n_x+n_y$ depicted in the inset of **Fig. 3D**. The slope of each linear fit was then plotted against the total number of animals $N$ (blue dots, **Fig. 3D**). The experimental data has a very close correspondence with the theoretical values in this plot (red line, **Fig. 3D**). For a high number of animals, both theory and data show Weber behavior, corresponding in this logarithmic plot to a straight line with slope -1 (black line, **Fig. 3D**) (24). Interestingly, for low numbers of animals, the theoretical prediction of a deviation from Weber´s behavior corresponds to the data.

The last case we consider has $k=1$, for which **Eq. (3)** depends only on the variable $\Delta N \equiv n_x - n_y$. This situation could take place when there is a high probability that only one of the options is good and those animals choosing *x* indicate that *x* may be the good one in a similar way that those choosing y may indicate that *x* might not be the good one. We have previously shown (22) that the simple decision rule $P_x = 1/(1 + a\,s^{-\Delta N})$ explains well a large data set of collective decisions in sticklebacks, *Gasterosteus aculeatus* (25, 26). In these experiments, animal groups were made to choose in two-choice set-ups with different combinations of social and non-social information (**Fig. 4A**, left). Interestingly, **Eq. (3)** has the simple rule $P_x = 1/(1 + a\,s^{-\Delta N})$ as a particular case for *k*=1 (**Supporting Text**). Indeed, all experimental results (blue histograms in **Fig. 4A** and **Fig. S7**) are fit using **Eq. (3)** with parameters *s*=2.5, *k*=1 (red lines in **Fig. 4A**). Additionally, for low numbers of animals (up to τ in **Fig. 1B**) an approximated $\Delta N$ rule can also be found for any value of *k* but with different values of the non-social reliability parameter *a* (**Supporting Text**). Therefore, the stickleback data can be fit with any value of *k* (green and blue lines in **Fig. 4A** and **Fig. S7** for *k*=0.5 and *k*=0, respectively), with robust fits (**Fig. S8**). The reason why in this case *k* can have any value is that its main effect is to control the slope of the boundaries of the plateau of probability 0.5, which is not present in the experimentally explored region of the stickleback dataset (white triangle, **Fig. 4B**). Still, all these fits have in common an effective $\Delta N$ rule for the experimental region (**Fig. 4B**), giving strong support to this rule in this dataset.

**Discussion**

Our results support that estimation by the brain using social information to counteract the ambiguity of sensory data is a fundamental principle in collective decision-making. The theory explains also the diversity in number discrimination schemes used in collective decisions, including 'counting up to a given number of animals', counting the difference of animals choosing among options, $\Delta N$, or the relative difference $\Delta N/N$, as well as observed deviations from these ideal cases and the existence of different counting regimes for high and low numbers as observed in many species, including humans (47, 48). A single mathematical rule contains all these cases and can be used as a first-principles approach to quantitatively study decisions in animal collectives.

One important ingredient of our theory is the use of probability matching, **Eq. (2)**. For symmetric decisions it implies a functional form of the type $P_x = f(x,y)/(f(x,y)+f(y,x))$. Our model in Eq. (3) is a particular case of this function, with $f(x,y)$ derived from an approximation to Bayesian estimation. Interestingly, many previous approaches derive from the form $P_x = f(x)/(f(x)+f(y))$ (21, 22, 27, 28), which is also a particular case of $P_x = f(x,y)/(f(x,y)+f(y,x))$, and therefore compatible with probability matching. In other cases, the basic form $P_x = f(x)/(f(x)+f(y))$ has been modified by adding constant terms (29, 30) or an extra function (25), as $P_x = f(x)/(f(x)+f(k))$ with $k$ a constant when animals have access to a single choice (31, 32, 34). Weber behavior can also be seen as a particular case. It has been previously described using a function (24) that can be expressed as $f(x,y) = 1/2 + \delta(n_x - n_y)/(n_x + n_y)$ with $\delta$ between 0 and 1/2. This function obeys $f(x,y) + f(y,x) = 1$ so in this case $P_x = f(x,y)$, following Weber's behavior.

These previous functions are very useful when applied to particular datasets as they may use few parameters in these conditions. In particular, our previous model (22), a particular case of **Eq. (3)** (**Supporting Text**), used only one parameter in the symmetric experiments with sticklebacks, and a model with two parameters described the ants dataset (24). However, these two models cannot fit the three datasets or even two of them (**Figures S9A,D and S10**). For the zebrafish data in **Fig. 2,** none of previously proposed functions (21, 22, 24, 27-29) give a good fit of the plateau in the data (**Fig. S9**). Our approach has been developed to be applied in very different species and conditions, here tested for three large datasets in three different species. One important factor in this ability to describe different datasets is that our basic function $f(x,y)$ has a term $s^{-(n_x - k n_y)}$ that captures how the estimated quality of an option depends not only on the animals choosing that option but also on the animals choosing the other option. These two sources of information are balanced by parameter *k*, and different datasets are found to correspond to different balances *k*.

Previous functions describing ant foraging include a constant term that represents a threshold of pheromone concentration below which ants do not react (24, 27, 28). In this way, these functions can describe the deviation from Weber's law at low pheromone concentration (24). In our case, the theory naturally shows this behavior as one more particular case of the predicted difference between low and high number of animals. Comparing the two approaches, it is interesting to consider that the behavior for low numbers that is predicted from estimation theory can be achieved in ants using a threshold of pheromone concentration.

An advantage of our approach is that the form of the function *f* is derived for any type of set-up simply from estimation given non-social sensory data and the behaviors of others (**Supplementary Text**). For example, we predict for a symmetric set-up with *N* options a generalization of **Eq. (1)** of the form

$$P(x \text{ is good}) = \frac{1}{1 + as^{-(n_x - kM)}},$$

with $M = \sum_{i \neq x}^{N} n_i$ the total number of animals choosing any option except *x*; see **Eq. (S10)** for the more general case of asymmetric choices.

A further advantage is that the parameters *a*, *s* and *k* are not only fitting parameters but have expressions, **Eqs. (S4)**, **(S9)** and **(S16)** respectively, that give additional predictions. For example, the social reliability parameter is given by

$$s = \frac{P(\beta | X, C)}{P(\beta | \overline{X}, C)},$$

with *β* a given animal behavior. This expression means that the social reliability parameter *s* is higher for a behavior *β* that is produced with high probability when *x* is a good option, and with very low probability when it is not a good option. Among all behaviors, those with higher *s* allow an individual to obtain a higher probability that option *x* is a good one, **Eq. (1)**, so we expect them to have a larger effect on collective decision-making.

Another advantage of an approach based on a theory of estimation is that generalizations of the theoretical expressions can be envisaged deriving models using fewer assumptions. For example, including dependencies in the behaviors of the other individuals and explicit space and time variables should be natural extensions of the theory.

**Materials and Methods**

**Experimental protocol for zebrafish**

All procedures met with the European guidelines for animal experiments (86/609/EEC). We used wild-type adult zebrafish, *Danio rerio*, of both sexes. Fish were acclimatized to the set-up water for one day before the experiments (**Fig. S11**). One hour before the experiment, each fish was isolated and fed to ensure uniform nutritional status across individuals. A focal fish entered the setup and swam freely in a central chamber between two 'social chambers' with different number of fish and separated from the choice chamber by glass. Once a fish had been recorded for 5 min it could be placed in one of the lateral chambers as a social stimulus for another fish. The fish in the lateral chambers were interchanged between trials to ensure uniformity, and sides were randomized. The central chamber of the set-up was washed between trials to remove odor traces. We computed the probability $P_x$ as the fraction of time the focal fish spent on the black region close to one of the social chambers, *x*. This fraction of time converges to $P_x$ for a fish that makes repeated decisions choosing *x* (*y*) with probability $P_x$ ($P_y=1-P_x$). A total of 238 fish were tested only once. In order to test the effect of previous experience, another 233 trials were performed with fish that were tested several times. We found no significant difference between the two groups in the mean times spent at each side (**Fig. S12**) so all data were pooled for **Fig. 2**.

**Model with noise added to the decision rule**

The model in **Eq. (3)** has a good agreement with data from experiments using the Argentine ant, *Linepithema humile* (24), except that experimental probabilities do not reach values as close to 0 or 1 as the theory. To account for the experimental data, we made a simple modification of the model by assuming that the ant has some probability $p_{rand}$ of making the decision at random motivated by unknown factors. Then, with probability (1-$p_{rand}$) the ant makes the decision according to **Eq. (3)**. Therefore, the probability of turning towards *x* is

$$P_x = \frac{p_{rand}}{2} + (1 - p_{rand})\left(1 + \frac{1 + as^{-(n_x - k n_y)}}{1 + as^{-(n_y - k n_x)}}\right)^{-1} \quad (4)$$

The parameters that best fit the ant data are *a*=2.5, *s*=1.07, *k*=0.53, $p_{rand}$=0.39. This same model can be applied to the zebrafish and stickleback datasets but in these cases the best fit is obtained for $p_{rand} \approx 0$, that corresponds to **Eq. (3)**.

**Analysis of the ants dataset**

Both the raw dataset and pre-processing routines were provided by the authors of reference (24). We used their data assuming no evaporation of pheromone (this assumption does not change the results significantly (24)). We calculated from the data the probability of turning right or left, not a continuous angle, to compare directly to our predicted probabilities. To reduce the noise in the experimental maps of **Fig. 3**, we symmetrized the data so that the probability shown at point ($n_x$,$n_y$) is obtained as ($P_x$($n_x$,$n_y$)+(1-$P_x$($n_y$,$n_x$)))/2.

Experimental data from ref. (24) measures a quantity that is proportional to the number of ants previously at the left/right of the deciding ant, not directly the numbers, so the number of ants ($n_x$, $n_y$) used in the plots relate to the actual number of ants that count for the decision, ($n_{x,\text{true}}$, $n_{y,\text{true}}$) as $n_x = \lambda\, n_{x,\text{true}}$, $n_y = \lambda\, n_{y,\text{true}}$, where $\lambda$ is an unknown proportionality constant. This relation means that the model still applies but with $s=s_{\text{true}}^{\lambda}$, where $s_{\text{true}}$ is the actual value of the reliability parameter.

**Fitting procedures**

In order to fit the model's parameters to the data, we performed 2-dimensional exhaustive searches in the space of parameters. For functions with more than 2 parameters, we performed the search successively with all possible pairs of parameters. In these cases we repeated the fit several times starting from different initial conditions, always getting the same final result.

**Acknowledgements:**

We are very grateful to Andrea Perna for sharing the raw ant data before publication. We thank David Sumpter, Jacques Gautrais, Raphael Jeanson, Matthieu Moreau and the members of de Polavieja Lab for fruitful discussions, and two reviewers for constructive comments. G.G.d.P. acknowledges funding by MICINN (Spain) as Plan Nacional and as partners of the ERASysBio+ initiative supported under the EU ERA-NET Plus scheme in FP7, including contracts to S.A and A.P.E. A.P.E. also acknowledges a FPU fellowship from MICINN (Spain).

**Contributions:**

S.A. designed project, performed experiments, analyzed data and performed modeling, A.P-E designed project, helped with experiments, analyzed data and performed modeling. G.G.d.P. designed and supervised project, performed modeling and wrote the paper with contributions from the other authors.

**Competing financial interests:**

The authors declare no competing financial interests.

**Figures:**

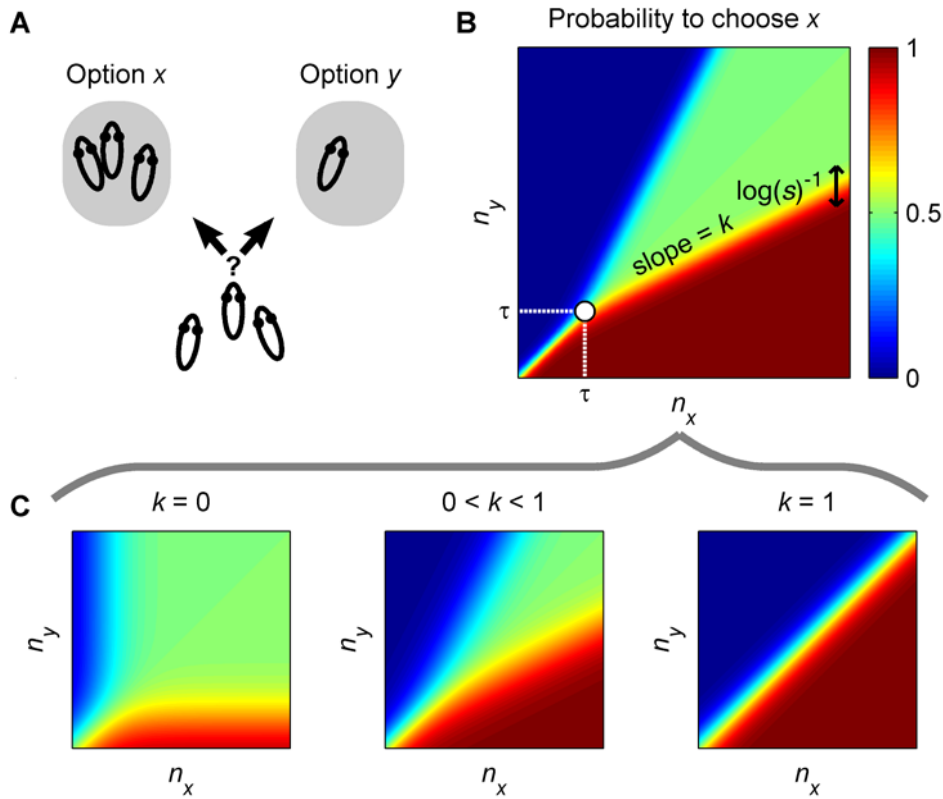

**Fig. 1. A general decision-making rule in animal collectives.** **(A)** Decision-making between two sites when $n_x$ and $n_y$ animals have already chosen sites $x$ and $y$, respectively. **(B)** The probability of choosing $x$ in the general rule ,**Eq. (3)**, plotted as a function of the animals that have already chosen between the two sites, $n_x$ and $n_y$. The theory predicts very different structure in the probability for the case of low and high number of animals, separated by point $\tau = \log(a)/(\log(s)(1-k))$. The rate of change of $P_x$ in the transition regions depends on the reliability parameter $s$, with the width of these regions proportional to $1/\log(s)$. **(C)** Same as **B** but for three different values of parameter $k$: $k=0$ (left), $0<k<1$ (middle) and $k=1$ (right).

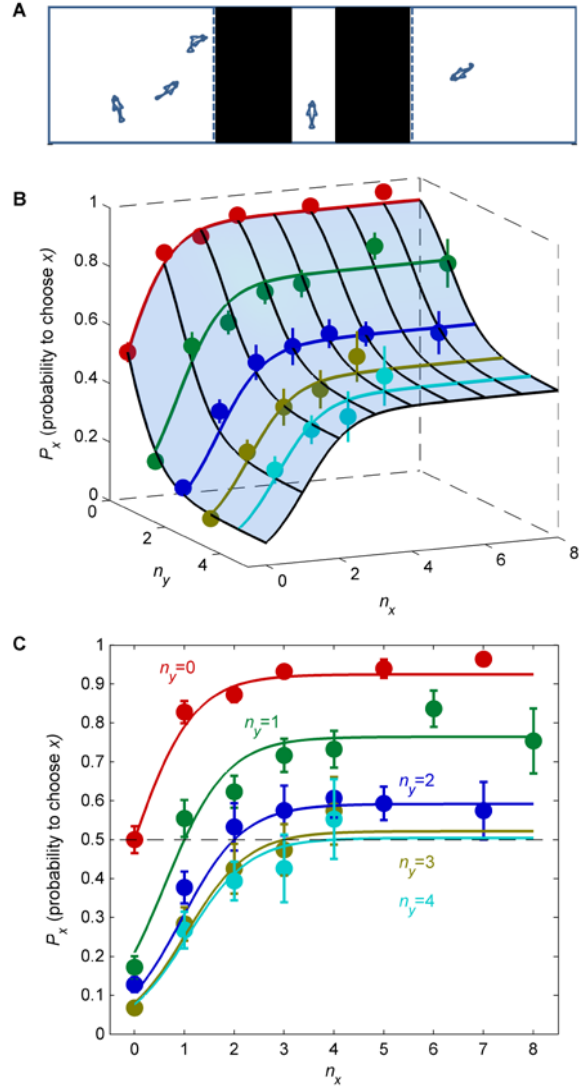

**Fig. 2. Zebrafish choices correspond to the general rule of decisions in collectives.**
(**A**) Focal fish choosing between two sites with different number of zebrafish, separated from the focal fish by glass. (**B**) Probability of choosing option *x* for different numbers of zebrafish at sites *x* and *y*, $n_x$ and $n_y$. Theoretical probabilities for $a=11.2$ and $s=5$ and $k=0$ in **Eq. (3)** represented as a surface and experimental data represented as dots indicating the mean value of typically 15 animals at each configuration. Different dot colors correspond to different values of $n_y$ and bars are SEM. (**C**) Same as **B** but plotted only as a function of $n_x$ and different colors representing the value of $n_y$.

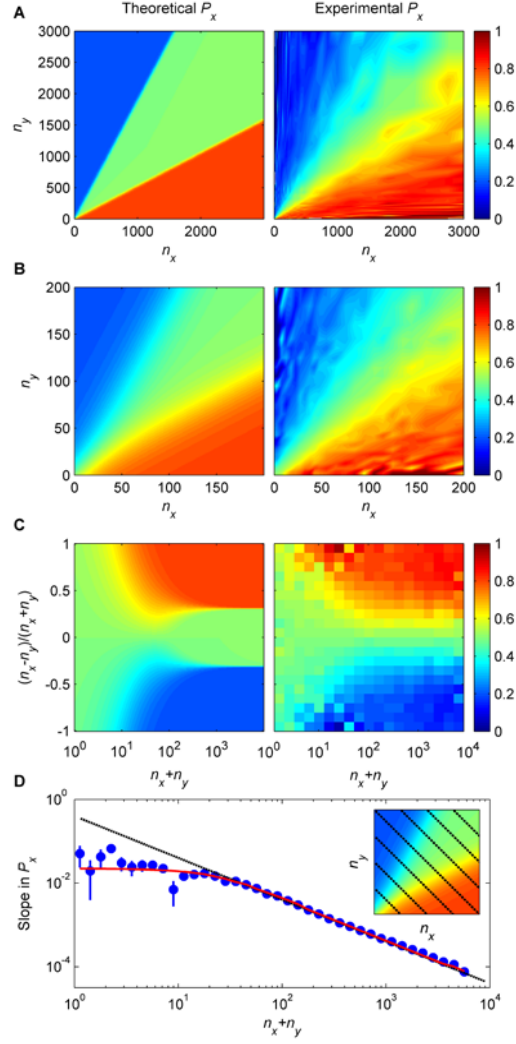

**Fig. 3. Ant choices correspond to the general rule of decisions in collectives.** (**A**) Probability of choosing option $x$ as a function of how many ants have previously been at locations $x$ and $y$, $n_x$ and $n_y$, for theory (left) using **Eq. (4)** with $a$=2.5, $s$=1.07, $k$=0.53, $p_{rand}$=0.39 and experiments (right) from ref. (24). (**B**) Detail of **A**. (**C**) Same as **A** but represented as a function of $\Delta N / N$ and $N$. (**D**) Slope of the probability of choosing $x$ in **A** as obtained from a linear fit along the lines depicted in the inset. Experimental values (blue dots; error bars are 95% confidence interval), theory (red line) and Weber law (black line).

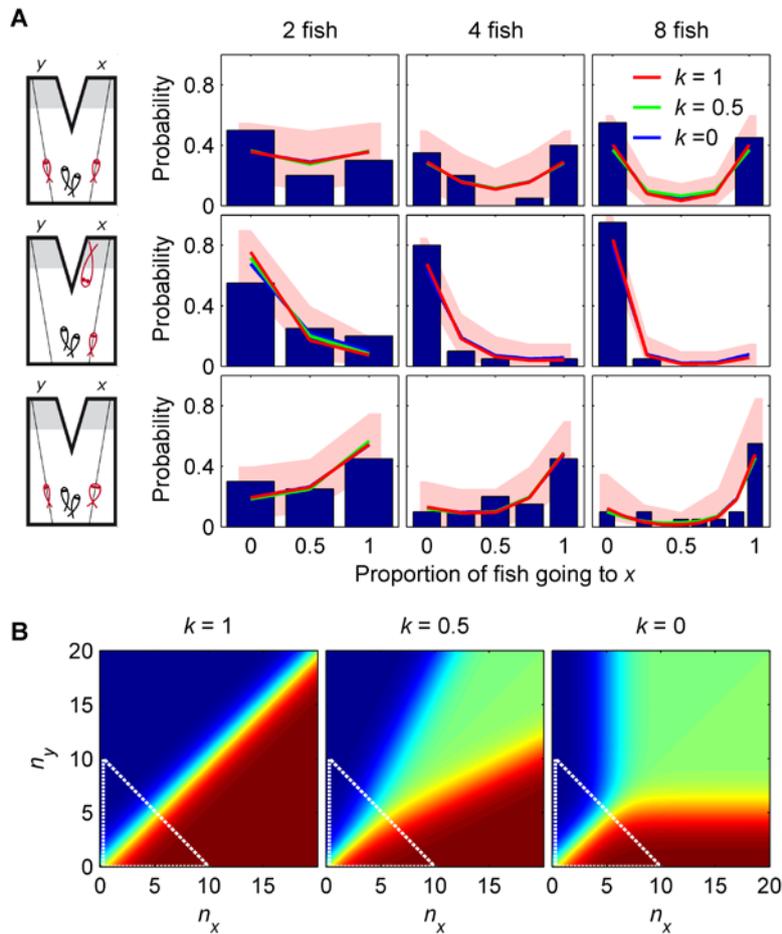

**Fig. 4. Stickleback choices correspond to the general rule of decisions in collectives. (A)** Probability of finding a final proportion of sticklebacks choosing option *x* (blue histograms are experimental results from refs. (25, 26) and theoretical values as lines for *k*=1, *k*=0.5 and *k*=0) for different group sizes (2, 4 and 8 fish) and for three types of set-ups: a symmetric set-up with different numbers of replica fish going to *x* and *y* (top), a set-up with a replica predator at *x* and different replica fish going to *x* (middle) and a symmetric set-up with modified replica fish (bottom). See model parameters and 68 additional experiments with fits in **Fig. S7**. **(B)** Theoretical $P_x$ for *k*=1, *a*=1 (left), *k*=0.5, *a*=5 (center) and *k*=0, *a*=224 (right) and *s*=2.5 in the three cases. All models require an effective Δ*N* rule to compare with data for the number of animals used in experiments (triangle).

**Supporting Information:**

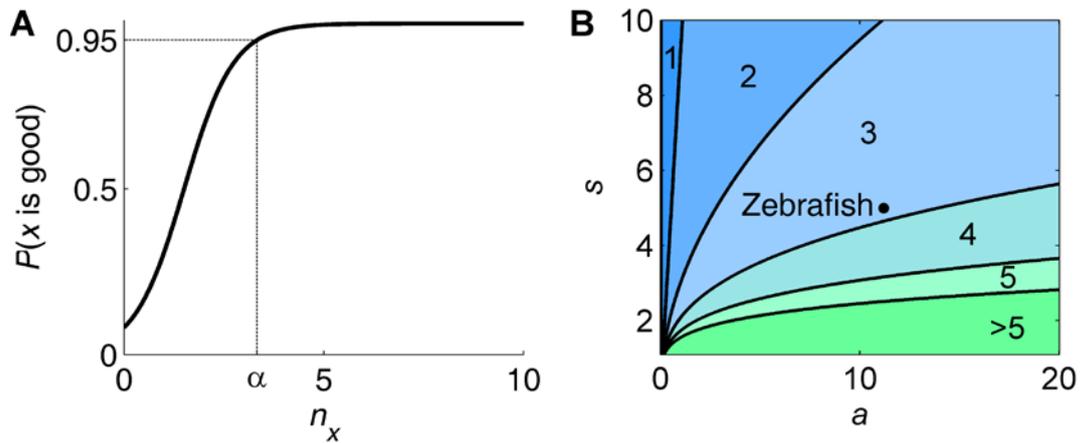

**Fig. S1: Maximum number of individuals ($\alpha$) that need to be counted according to the model for *k*=0. (A)** Probability that option *x* is good, **Eq. (1)**, here plotted for parameters *a*=11.2, *s*=5 and *k*=0. For *k*=0 this probability only depends on the variable $n_x$, increasing as $n_x$ increases until a value of 1. We compute $\alpha$ as the value of $n_x$ for which the probability in **Eq. (1)** reaches 0.95, getting $\alpha = (\log(a) - \log(1/0.95 - 1))/\log(s)$. Since for *k*=0 the probability to choose *x*, Eq. (3), only depends on $n_x$ through *P(x* is good), to make the decision the animals do not need to keep count of $n_x$ beyond $\alpha$. **(B)** Number up to which animals need to count, $\alpha$, as a function of parameters *a* and *s*. For the parameters of the zebrafish dataset, animals only need to count up to 3.

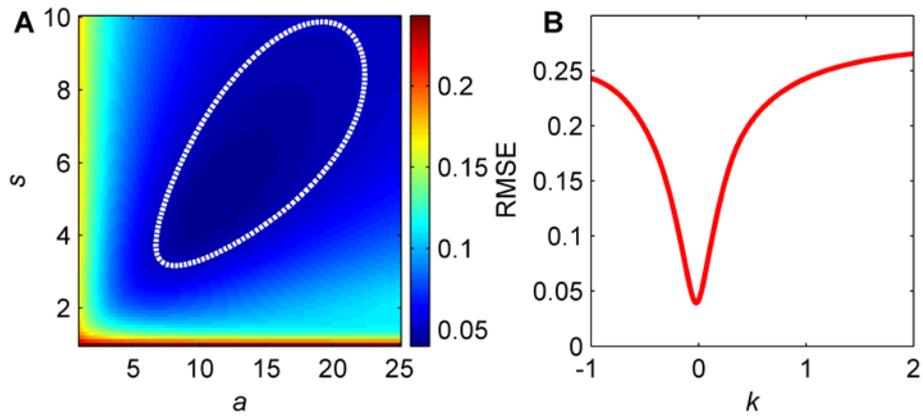

**Fig. S2: Robustness of fit to zebrafish data. (a)** Root mean squared error between model predictions and data as a function of *a* and *s* (*k*=0). The dotted line limits the region with error below 0.05. **(b)** Root mean squared error between model and data as a function of *k* (*a*=11.2, *s*=5).

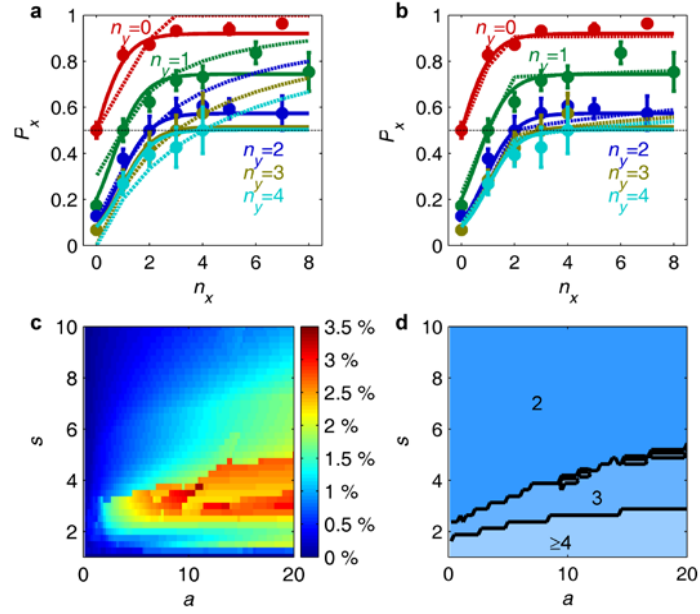

**Fig. S3: A simple mechanistic model gives an approximation to Eq. (3). a.** Comparison between decisions using a simple stochastic model (dashed lines) and the model in Eq. (3) of the main text (solid lines). In the stochastic model the focal fish either follows one of the other fish present in the setup (going to the zone where the followed fish is) or does not follow anyone (and therefore moves randomly). If there are $N$ fish in the set-up (apart from the focal one), the focal fish will follow any of them with equal probability $P$ when $NP<1$ and otherwise with probability $1/N$. The probability of not following another fish, and thus choosing at random, is then max($\{1-NP, 0\}$). We modeled the experiment as a series of repeated decisions following this rule, and calculated the time spent at each side in the limit of infinite decisions. Despite the simplicity of this simple stochastic model, it already shows some of the qualitative features of the data. **b.** Same as **a** but now the stochastic model considers that the focal fish has a different probability to follow close and far individuals. The implementation of the model was as follows. The probability of not following anyone is now max($\{(1-N_{close}P_{close}-N_{far}P_{far}),P_{nF}\}$), where $N_{close}$ ($N_{far}$) is the number of fish in the same (opposite) zone as the focal fish, and $P_{nf}$ is the minimum probability of not following anyone. When $N_{close}P_{close}+N_{far}P_{far}>1-P_{nF}$, $P_{close}$ and $P_{far}$ are renormalized so that $N_{close}P_{close}+N_{far}P_{far}=1-P_{nF}$ while $P_{close}/P_{far}$ remains constant. The model with $P_{close}=0.71$, $P_{far}=0.005$, $P_{nF}=0.1$ (dashed lines), has a very good agreement both with the model in Eq. (3) (solid lines) and the experimental data (points). **c.** Difference between the model in Eq. (3) and the mechanistic model in **b** as a function of $a$ and $s$. For most values, there is a close agreement. **d.** Maximum number of individuals that is necessary to count according to model in **b** when parameters are fitted to match the model in Eq. (3). For most parameter values, we can make $P_{far}=0$ without a significant worsening of the fit. Then, the probability of not following any fish is max$\{(1-N_{close}P_{close}), P_{nF}\}$, that saturates when $N_{close} \geq (1-P_{nF})/P_{close}$. Due to this saturation, the fish only needs to count up to $(1-P_{nF})/P_{close}$. This model is consistent with the notion that for a very wide parameter range animals only need to count up to a small number.

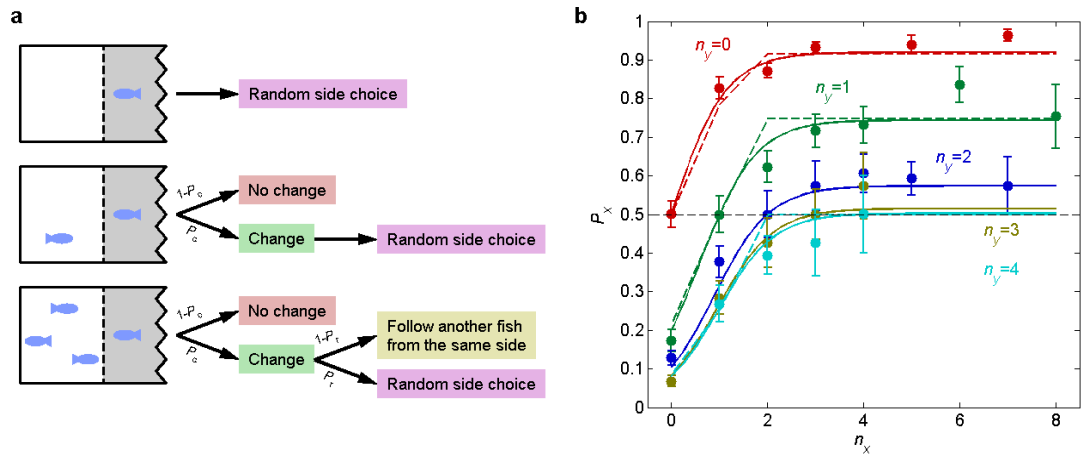

**Fig. S4: A very simple mechanistic model gives an approximation to Eq. (3) for parameters corresponding to zebrafish data. a**. Schematic diagram of the model. The focal fish (the one on the shaded area) only takes into account the fish that are at the same side. If there are no other fish at the same side, the focal fish moves randomly, and therefore has probability ½ of choosing any side at the next time step (top). If there are other fish at its side, the focal fish follows one of them. At the next decision, it chooses either to stay following the same fish (with probability 1-$P_c$) or to change (with probability $P_c$) and follow another fish, or not follow anyone. If there is only one fish at the same side, changing means necessarily not following anyone in the next time step, and therefore moving randomly (middle). If there are more than one fish, then changing may lead to follow another fish and therefore remain at the same side, with probability 1-$P_r$, or not follow anyone, with probability $P_r$ (bottom). **b.** Comparison between model in **a** (dashed lines), model in Eq. (3) (solid lines) and experimental data (points) for $P_c$=0.28 and $P_r$=0.34. The correspondence is good except for the $n_y$=2 case (blue). The model corresponds to "counting up to 2", while the data is best fitted with a "counting up to 3" model as in the more complex model of **Fig. S3b**.

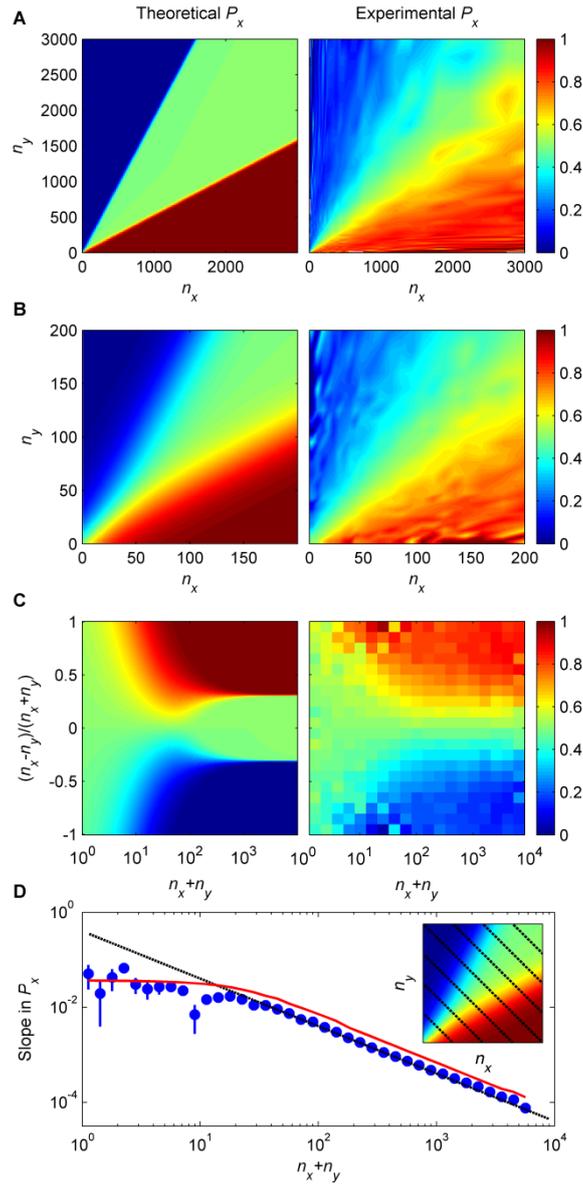

**Figure S5. Eq. (3) reproduces the structure of the ants dataset.** Same as Fig. 3 in main text but comparing the ants dataset to Eq. (3) (or, equivalently, Eq. (4) setting $p_\text{rand}$=0 instead of the value $p_\text{rand}$=0.39 in main text)

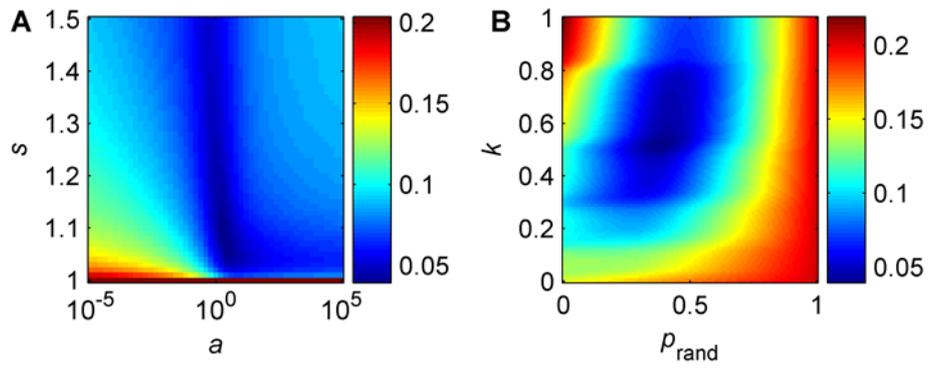

**Fig. S6: Robustness of the fit of the model in Eq. (4) to the ant dataset. a.** Mean squared error between model and data as a function of parameters *a* and *s*, for $k=0.53$ and $p_{rand}=0.39$. In order to adequately sample the data, that span several orders of magnitude, we scanned the $n_x$-$n_y$ plane using sections of constant $n_x+n_y$ equispaced in a logarithmic scale, instead of a square grid. **b.** Mean squared error as a function of *k* and $p_{rand}$, for $a=2.5$ and $s=1.07$. Sampling of the $n_x$-$n_y$ plane as for **a**.

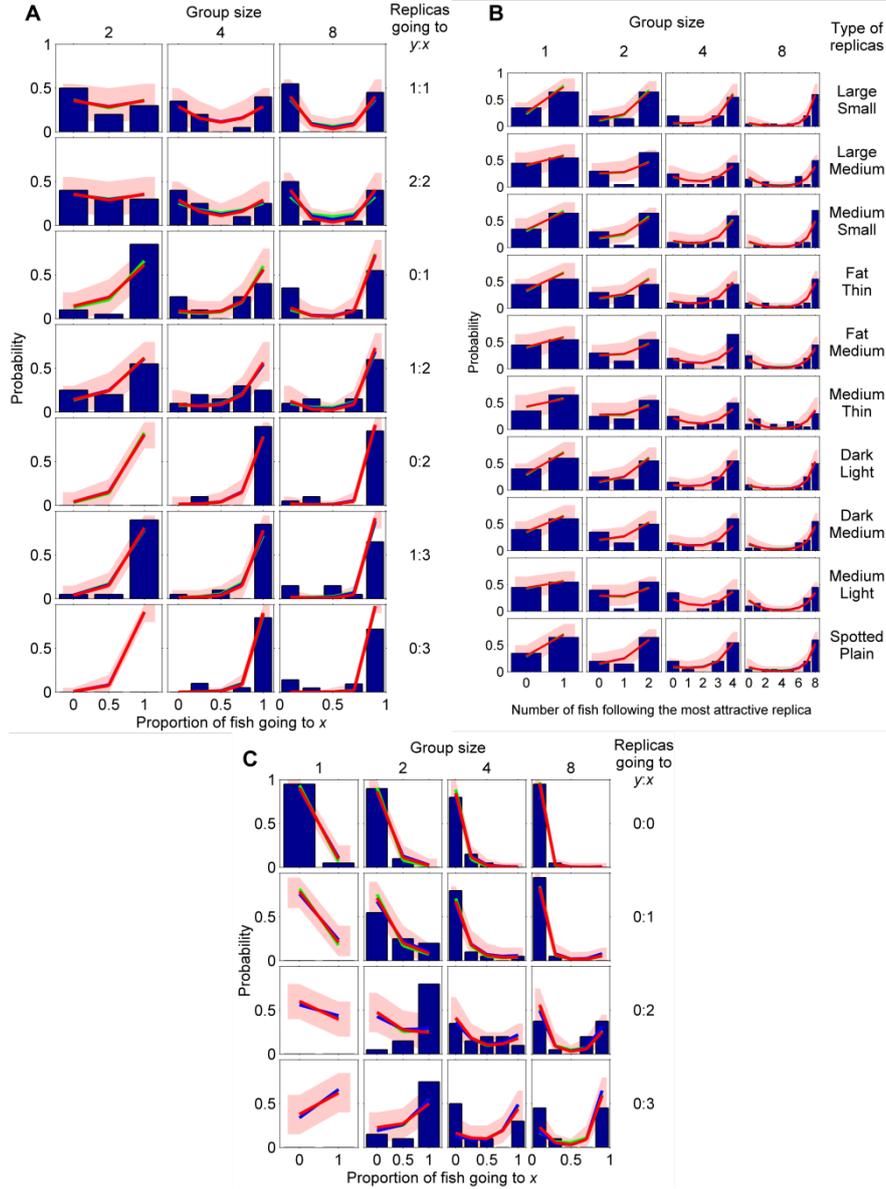

**Fig. S7: Complete stickleback dataset (25, 26) and model fits.** The three figures show experimental data as blue histograms and results for the $k=1$ model (22) as red lines and green and blue lines for $k=0.5$ and $k=0$, respectively. In the three cases $s=2.5$ and $a$ was refitted for each $k$. Pink regions limit the 95% confidence intervals for the $k=1$ case. **a.** Results for symmetric set-up with different number of replica fish going to each side (for example, 1:2 means one replica going to $y$ and 2 replicas going to $x$). $a_x=a_y=1$ for $k=1$ (red line), $a_x=a_y=5$ for $k=0.5$ (green line) and $a_x=a_y=224$ for $k=0$ (blue line). **b.** Results for symmetric set-up and differently modified replica fish going to each side. We set the intermediate replica's reliability parameter equal to the one of the real fish ($s=2.5$), and adjust the others to match the ratios found in Ref. (22). We got $s_{small}=1.25$, $s_{medium}=2.5$ $s_{large}=3.57$, $s_{thin}=1.88$, $s_{medium}=2.5$ $s_{fat}=3.62$ $s_{light}=1.95$, $s_{medium}=2.5$ $s_{dark}=4.55$, $s_{plain}=2.5$ $s_{spotted}=5.81$. Parameter $a$ as in **a**. **c.** Results for set-up with a replica predator at $x$. $a_x=9.5$, $a_y=1/9.5$ for $k=1$ (red line), $a_x=1.25$, $a_y=31.5$ for $k=0.5$ (green line) and $a_x=1250$, $a_y=10000$ (in this case if we multiply these two parameters by any number greater than 0.1, the fit changes very little) for $k=0$ (blue line).

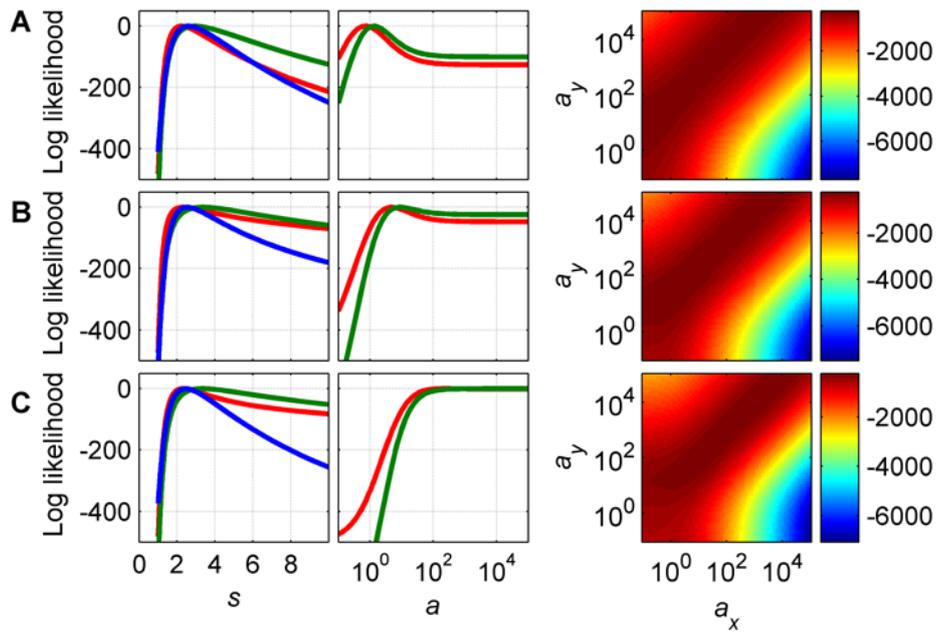

**Fig. S8: Robustness of the fit for the stickleback dataset. a.** Results of fits for $k$=1. Left: Log-likelihood as a function of parameter $s$ (symmetric set-up in red, set-up with two different types of replicas in green, setup with predator in blue). Middle: Log-likelihood as a function of parameter $a$ (red for symmetric set-up and green for set-up with modified replicas). Right: Log-likelihood as a function of parameters $a_x$ and $a_y$ for the asymmetric set-up with predator. **b.** Same as **a** but for $k$=0.5. **c.** Same as **a** but for $k$=0. All log-likelihoods are relative to their maximum value.

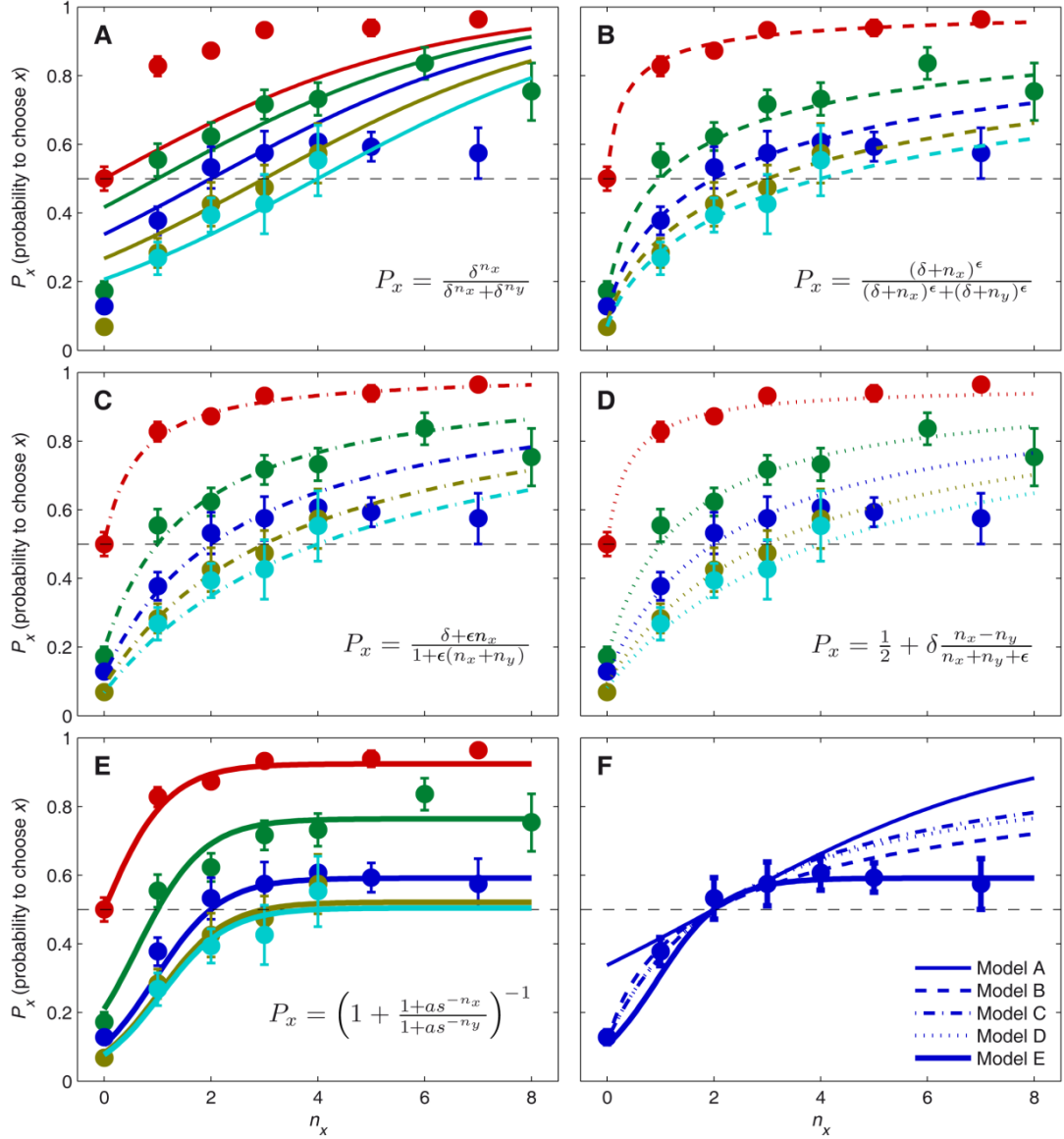

**Figure S9. Best fit of different functions to zebrafish dataset. (A)** Logistic function $P_x = \delta^{n_x}/(\delta^{n_x} + \delta^{n_y})$, as in refs. (21, 22), for $\delta = 1.4$. **(B)** $P_x = (\delta + n_x)^\varepsilon / ((\delta + n_x)^\varepsilon + (\delta + n_y)^\varepsilon)$, as in refs. (27, 28), for $\delta = 0.1$ and $\varepsilon = 0.7$. **(C)** $P_x = (\delta + \varepsilon n_x)/(1 + \varepsilon(n_x + n_y))$, as in ref. (29), for $\delta = 0.5$ and $\varepsilon = 1.6$. **(D)** $P_x = 0.5 + \delta(n_x - n_y)/(n_x + n_y + \varepsilon)$, as in ref. (24), for $\delta = 0.48$ and $\varepsilon = 0.47$. **(E)** Our model in **Eq. (3)** of main text, for $a$=11.2, $s$=5. **(F)** Comparison of the five previous models for line $n_y$=2. Only the model in **Eq. (3)** gives a good fit in this region.

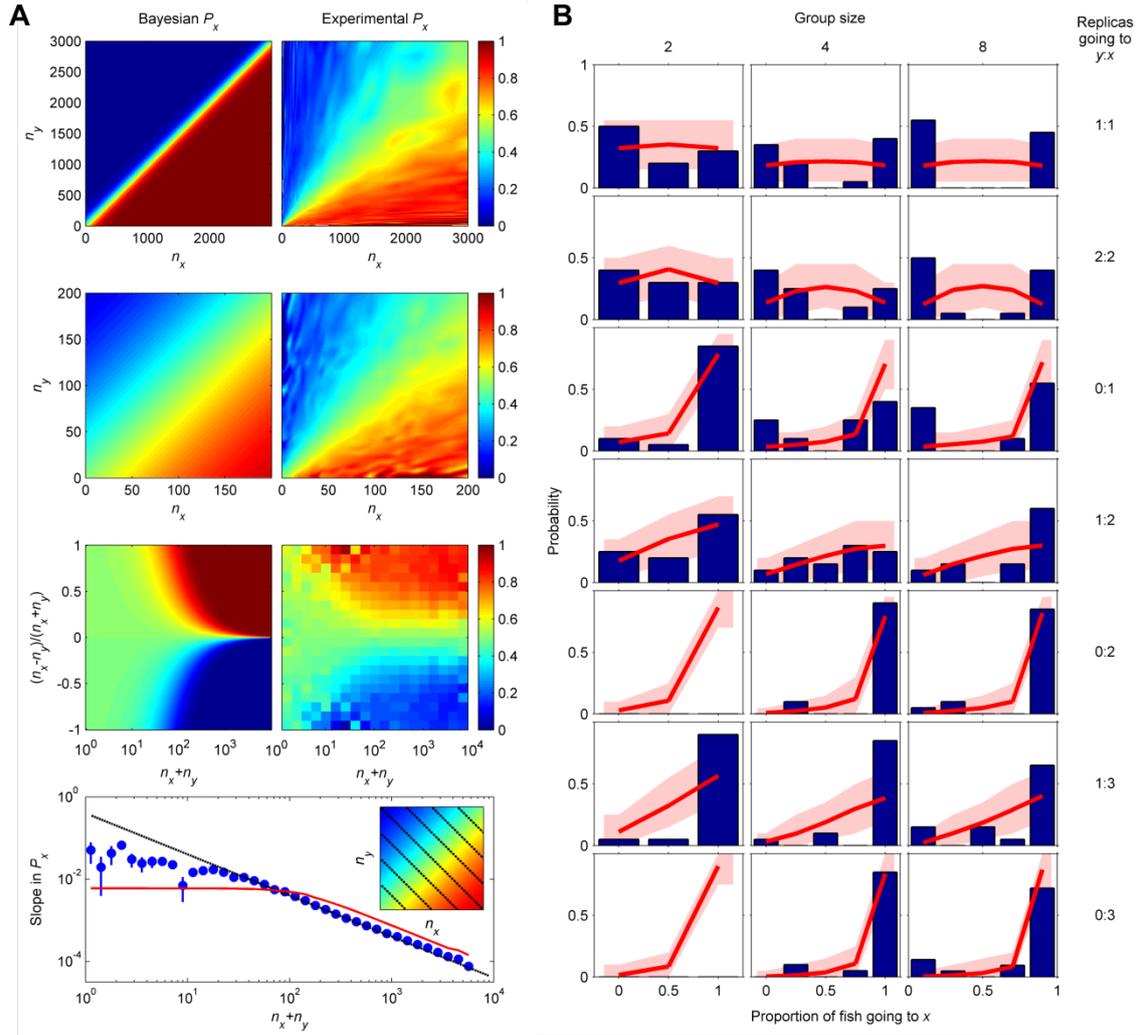

**Figure S10. The models in references (22, 24) do not explain other datasets. (A)** Same as **Figure 3** in main text, but using $P_x = \left(1 + s^{-(n_x - n_y)}\right)^{-1}$, with $s$=1.012. This model was used in (22) to describe the stickleback dataset, and cannot describe the ants dataset. **(B)** Same as **Figure S7A**, but with $P_x = 0.5 + A(n_x - n_y)/(n_x + n_y + T)$, with $A$=0.5 and $T$=0.4. This is the function used in (24) to describe the ant dataset, with the 0.5 term added and with $A$ restricted between 0 and 0.5, so that probabilities are between 0 and 1.

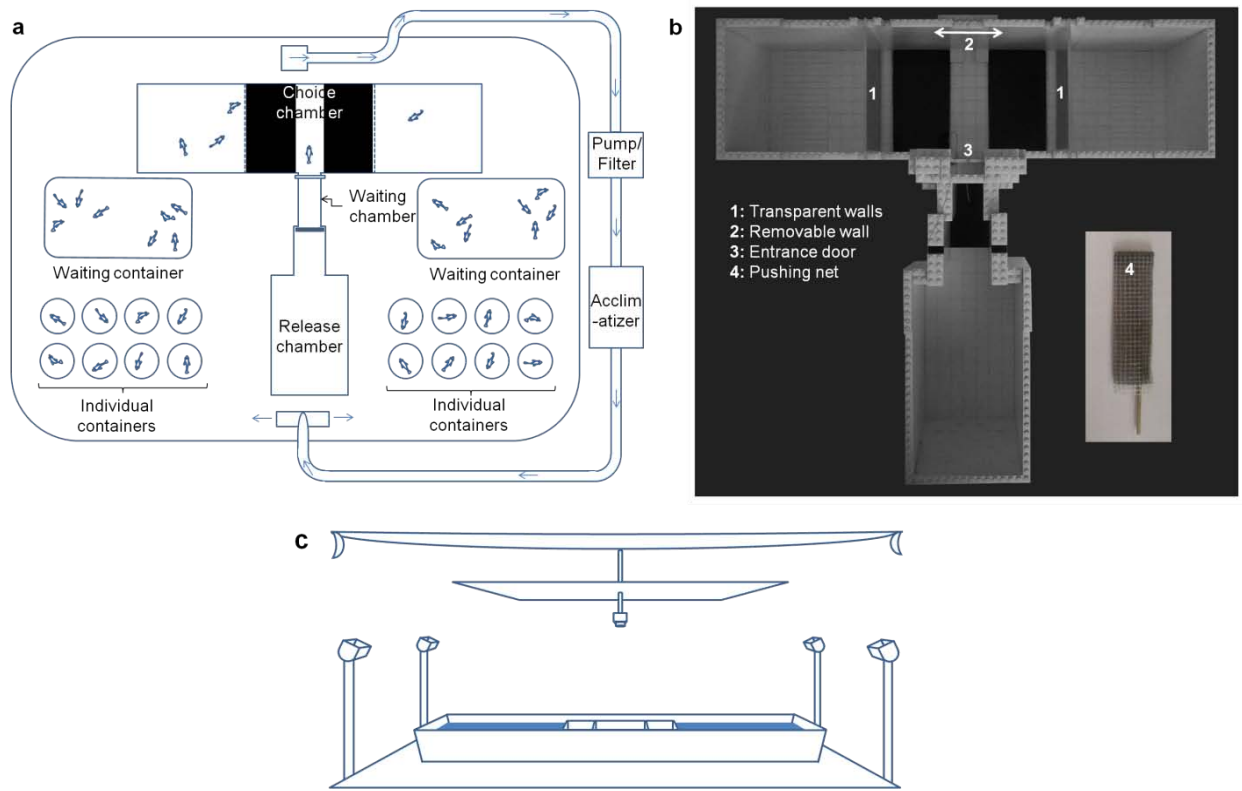

**Fig. S11: Experimental setup for zebrafish. a.** The behavioral setup is inside a bigger tank so that fish are acclimatized to the same water for one day before the experiment, housed in waiting containers in groups of 8-10 fish. One hour before the experiment, each fish is isolated, and fed with frozen artemia in an individual container. The fish stays in the individual container until placed in the release chamber and gently pushed into the waiting chamber with a net that fits tightly between the walls to prevent the fish from going back to the release chamber. The door to the set-up is then lifted and, once the fish enters the setup, it is closed. The camera records for 5 minutes from the opening of the door. After the experiment, the fish is pushed back to the release chamber, where it is caught. Then, a segment of wall opposite to the entrance door is removed, and water from outside is pumped into the central chamber so that odors are washed out. **b.** The T-shaped set-up is made of white LEGO[TM] bricks, with transparent walls separating the three chambers made of UV-transparent plexiglass (PLEXIGLAS GS 2458, Evonik Para-Chemie GmbH, Gramatneusiedel, Austria). The set-up's central chamber (choice chamber) measures 20x13 cm. The floor of this central chamber has a central white zone of 5 cm wide, and two black lateral zones of 7.5 cm wide each. The two lateral chambers measure 14x13 cm each. Walls are 17 cm high but water level was 6 cm. **c.** Illumination is provided by four 500W halogen lamps pointing to a white sheet on the ceiling. A Basler A622f camera records from above. An opaque roof just above the camera provides uniform shading on the set-up.

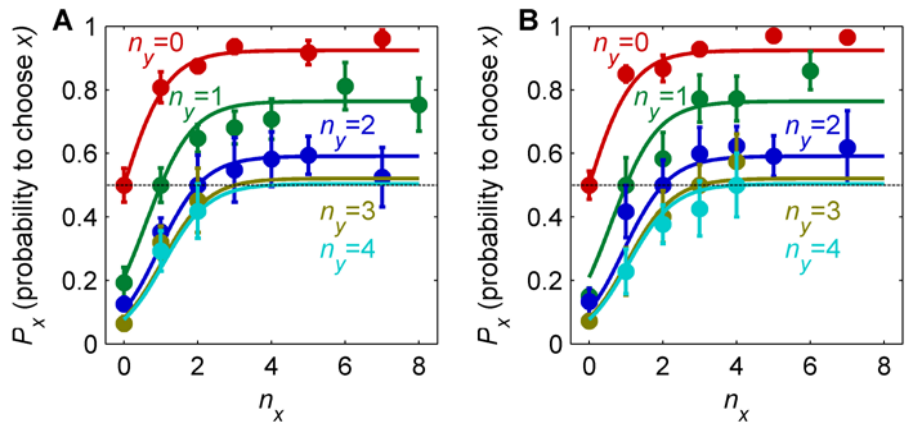

**Fig. S12: Comparison of results using naive and non-naive zebrafish. (a)** Results for naive zebrafish, which have never seen the set-up before the experiment. **(b)** Results for zebrafish that have been tested several times in the setup. Lines correspond to the theoretical model, **Eq. (3)**, with same parameter values as for **Fig. 2C** in main text.

**Supporting Text**

Here we give (A) the derivation of Eq. (1) in main text, (B) derivation of a more general equation for an asymmetric set-up (for Figure 4A (middle) in main text), (C) of an equation for a symmetric set-up but with different types of animals to follow, as in Figure 4A (bottom) of main text. Additionally, (D) shows that the model in our ref. (22) is a particular case of Eq. (3). (E) Derivation of an expression for the point $\tau$ separating the low-number and high-number decision behaviours, and proof of the approximate $\Delta N$ rule for low $N$.

### (A) Derivation of Eq. (1) in main text

The following derivation follows similar steps to our derivation in (22), except for the key difference that animals now estimate the probability that different options are good instead of the best. This simple difference makes the theory more general with previous results in (22) only a particular case, as shown in section (D).

Consider a focal individual making a decision among several options ($x$, $y$, $z$,...). To make this decision, it estimates the probability that each option is a good choice. `Good' may refer to presence of food, shelter, absence of predators, or any other feature. To perform this estimation it uses the information of the environment gathered directly by its sensors (non-social information, $C$), and the behaviours of the other individuals (social information, $B$). The probability that a given option (say, option $x$) is a good choice, given both non-social and social information is

$$P(X|C,B), \tag{S1}$$

where $X$ stands for `$x$ is a good choice'. We can compute this probability using Bayes' theorem,

$$P(X|C,B) = \frac{P(B|X,C)P(X|C)}{P(B|X,C)P(X|C) + P(B|\overline{X},C)P(\overline{X}|C)}, \tag{S2}$$

where $\overline{X}$ stands for '$x$ is not a good choice'. Dividing the numerator and denominator of Eq. (S2) by the numerator, we get

$$P(X|C,B) = \frac{1}{1 + a_x S_x}, \quad (S3)$$

with

$$a_x = \frac{P(\overline{X}|C)}{P(X|C)} \quad (S4)$$

and

$$S_x = \frac{P(B|\overline{X},C)}{P(B|X,C)}, \quad (S5)$$

where we use the subindex $x$ to indicate that it refers to the estimation for option $x$. Each of the options has a set of equations like (S3)-(S5). Note that $a_x$ only contains non-social information ($C$), so we call it non-social term, while the social information ($B$) is contained in the social term, $S_x$. A practical version of Eq. (S3) is obtained using the approximation that the focal individual does not take into account the correlations among the rest of individuals (however, see our ref. (22) for a treatment of these correlations). This assumption implies that the probability of a given set of behaviours is equal to the product of the probabilities of individual behaviours. We apply this to the probabilities needed to compute $S_x$ in Eq. (S5),

$$P(B|X,C) = Z \prod_{i=1}^{N} P(b_i|X,C), \quad (S6)$$

where $B$ is the set of behaviours of the other $N$ animals at the time the focal individual is choosing, $B = \{b_i\}_{i=1}^{N}$, and $b_i$ denotes the behaviour of individual $i$. $Z$ is a combinatorial term counting the number of possible decision sequences leading to the set of behaviours $B$, that will cancel out below. Substituting Eq. (S6), and an analogous expression for $P(B|\overline{X},C)$, into Eq. (S5), we get

$$S_x = \prod_{i=1}^{N} \frac{P(b_i|\overline{X},C)}{P(b_i|X,C)}. \qquad (S7)$$

A more useful expression is obtained if we consider, instead of the full individual behaviours ($b_i$) with all their details, a set of behavioural classes that group together the behaviours that contain similar information about the choice. For example, in a two-choice set-up, useful behavioural classes might be 'choosing $x$' (denoted as $\beta_x$) and 'choosing $y$' ($\beta_y$). Consider in general $L$ behavioural classes, $\{\beta_j\}_{j=1}^{L}$. We do not here consider animals to have individual differences, so all have the same probabilities for each behaviour, for example the same $P(\beta_1|X,C)$ and $P(\beta_1|\overline{X},C)$ for behaviour $\beta_1$. This means, for example, that if the first $n_1$ individuals are performing behaviour $\beta_1$, we have $\prod_{i=1}^{n_1} \frac{P(b_i|\overline{X},C)}{P(b_i|X,C)} = \left(\frac{P(\beta_1|\overline{X},C)}{P(\beta_1|X,C)}\right)^{n_1}$. We can then write Eq. (S7) as

$$S_x = \prod_{j=1}^{L} s_{xj}^{-n_j} \qquad (S8)$$

where $n_j$ is the number of individuals performing behaviour $\beta_j$, and

$$s_{xj} = \frac{P(\beta_j|X,C)}{P(\beta_j|\overline{X},C)}. \qquad (S9)$$

To summarize, the probability that option $x$ is a good choice is, using Eqs. (S3) and (S8),

$$P(X|C,B) = \left(1 + a_x \prod_{j=1}^{L} s_{xj}^{-n_j}\right)^{-1}, \qquad (S10)$$

with $a_x$ in Eq. (S4) and $s_{xj}$ in Eq. (S9).

The zebrafish experiments in the main text were performed in a set-up with two identical sites to choose from, except for the number of animals at each site, $n_x$ and $n_y$. The focal animal can observe two types of behaviours, 'stay at $x$' ($\beta_x$) and 'stay at $y$' ($\beta_y$). Eq. (S10) then reduces to

$$P(X|C,B) = \frac{1}{1 + a_x s_{xx}^{-n_x} s_{xy}^{-n_y}}. \tag{S11}$$

Similarly, for option *y* the estimation is

$$P(Y|C,B) = \frac{1}{1 + a_y s_{yy}^{-n_y} s_{yx}^{-n_x}}. \tag{S12}$$

The non-social information for the two sites *x* and *y* is identical by experimental design so

$$\begin{aligned} P(X|C) &= P(Y|C) \\ P(\bar{X}|C) &= P(\bar{Y}|C) \end{aligned}. \tag{S13}$$

These relations in Eq. (S13) mean that $a_x=a_y$, as it is clear from its definition in Eq. (S4). For notational simplicity we then define

$$a \equiv a_x = a_y. \tag{S14}$$

The symmetry of the set-up also implies the following relations

$$\begin{aligned} P(\beta_x|X,C) &= P(\beta_y|Y,C) \\ P(\beta_x|\bar{X},C) &= P(\beta_y|\bar{Y},C) \\ P(\beta_x|Y,C) &= P(\beta_y|X,C) \\ P(\beta_x|\bar{Y},C) &= P(\beta_y|\bar{X},C) \end{aligned} \tag{S15}$$

In an idealized situation in which the only possible behaviors were 'stay at *x*' and 'stay at *y*', we would have that $P(\beta_x|X,C) = 1 - P(\beta_y|X,C)$. As real behaviors are much more complex, and different behaviors can exist, these two probabilities will not sum one in general.

According to (S15) and (S9), we have that $s_{xx}=s_{yy}$ and $s_{xy}=s_{yx}$. It is then useful to define

$$\begin{aligned} s &\equiv s_{xx} = s_{yy} \\ k &\equiv -\frac{\log(s_{xy})}{\log(s_{xx})} = -\frac{\log(s_{yx})}{\log(s_{yy})} \end{aligned}. \tag{S16}$$

Using Eqs. (S14) and (S16), we can write Eqs. (S11) and (S12) as

$$P(X|C,B) = \frac{1}{1+as^{-(n_x-kn_y)}}$$
$$P(Y|C,B) = \frac{1}{1+as^{-(n_y-kn_x)}},$$  (S17)

obtaining Eq. (1) in the main text. Note that $s = P(\beta_x|X,C)/P(\beta_x|\overline{X},C) = P(\beta_y|Y,C)/P(\beta_y|\overline{Y},C)$, that is, the probability of choosing one option when it is a good choice over the probability of choosing it when it is a bad choice. Therefore, parameter $s$ measures how reliable are the choices of each of the other individuals.

The probability of choosing $x$ or $y$ is then obtained using probability matching, Eq. (2) in main text, to give Eq. (3) in main text,

$$P_x = \left(1 + \frac{1+as^{-(n_x-kn_y)}}{1+as^{-(n_y-kn_x)}}\right)^{-1}.$$  (S18)

**(B) Derivation of a more general equation for an asymmetric set-up, as in Figure 4A (middle) of the main text**

In the case of an asymmetric set-up, the non-social information for the two sites $x$ and $y$ is different so

$$P(X|C) \neq P(Y|C)$$
$$P(\overline{X}|C) \neq P(\overline{Y}|C).$$  (S19)

These relations in Eq. (S19) mean that $a_x \neq a_y$, as it is clear from its definition in Eq. (S4).

In the symmetric case we used the relations in Eq. (S15),

$$P(\beta_x|X,C) = P(\beta_y|Y,C)$$
$$P(\beta_x|\overline{X},C) = P(\beta_y|\overline{Y},C)$$
$$P(\beta_x|Y,C) = P(\beta_y|X,C)$$
$$P(\beta_x|\overline{Y},C) = P(\beta_y|\overline{X},C).$$
(S20)

As the non-social asymmetry can modulate the probabilities for the behaviours, these relations need not be satisfied exactly. However, this effect is probably much weaker than the effect of the non-social asymmetry on the non-social term in Eq. (S19). Therefore, for simplicity we use relations (S20) also for the asymmetric setup. The good fit with experimental data confirms that they are a good approximation.

According to (S20) and (S9), we have that $s_{xx}=s_{yy}$ and $s_{xy}=s_{yx}$ and using the definitions in Eq. (S16), we find that Eq. (S11) and (S12) become

$$P(X|C,B) = \frac{1}{1+a_x s^{-(n_x-kn_y)}}$$
$$P(Y|C,B) = \frac{1}{1+a_y s^{-(n_y-kn_x)}}.$$
(S21)

The probability of choosing $x$ or $y$ is then obtained using probability matching, Eq. (2) in main text, to get

$$P_x = \left(1 + \frac{1+a_x s^{-(n_x-kn_y)}}{1+a_y s^{-(n_y-kn_x)}}\right)^{-1},$$
(S22)

represented in **Fig. S13**.

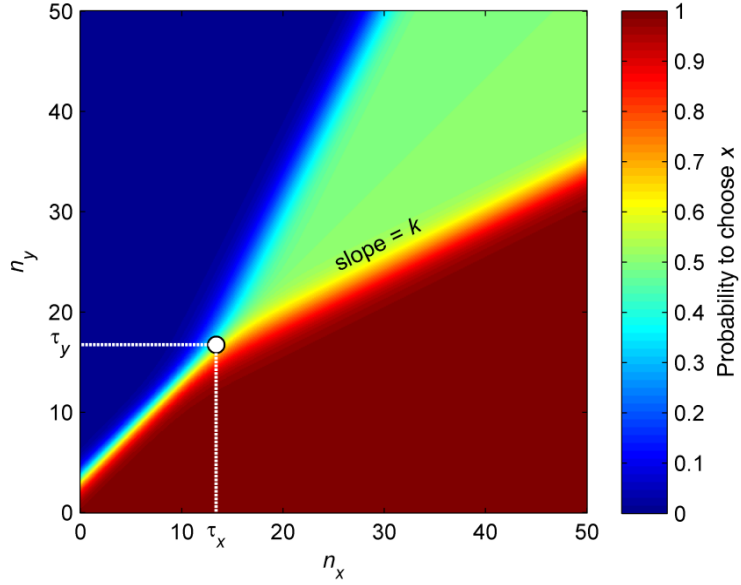

**Figure S13. Probability to choose *x*, $P_x$, for the general case of asymmetric non-social information, Eq. (S22).** Parameter: for *s*=2.5, *k*=0.5, $a_x$=100, $a_y$=10000. Compare this figure with the one corresponding to symmetric social information in **Fig. 1B**. See Equation (S29) for an analytical expression of ($\tau_x$, $\tau_y$).

### (C) Derivation of an equation for a symmetric set-up but with different types of animals to follow, as in Figure 4A (bottom) of the main text

When there are different types of animals to follow, as in the third row of **Fig. 4A** of the main text, following the steps of section (A) we find that each type of animal has its own reliability *s*. For the particular case of the experiment of ref. (26), we have three different types of animals (real animals, the most attractive replica and the less attractive replica, with reliability parameters *s*, $s_R$ and $s_r$, respectively). When the most attractive replica goes to *x* and the less attractive one goes to *y*, Eq. (S18) becomes

$$P_x = \left(1 + \frac{1 + a\ s^{-(n_x - k n_y)} s_R^{-1} s_r^{k}}{1 + a\ s^{-(n_y - k n_x)} s_R^{k} s_r^{-1}}\right)^{-1}. \tag{S23}$$

### (D) Demonstration that the model in our ref. (22) is a particular case of Eq. (3)

The decision-making model we used in reference (22) was developed for a case in which an animal has to choose using the probability that an option is the best one, whereas the model in this paper is for estimated good options. In reference (22), we obtained that the probability of choosing x in a two choice set-up that can present an asymmetry as

$$P_x = \left(1 + a_{old}\, s^{-(n_x - n_y)}\right)^{-1}, \quad \text{(S24)}$$

with $a_{old}=1$ for the symmetric case.

Multiplying and dividing inside the brackets of Eq. (S24) by $\left(1 + \frac{1}{a_{old}} s^{-(n_y - n_x)}\right)$, we rewrite this expression as

$$P_x = \left(1 + \frac{1 + a_{old}\, s^{-(n_x - n_y)}}{1 + a_{old}^{-1} s^{-(n_y - n_x)}}\right)^{-1}, \quad \text{(S25)}$$

so Eq. (S22) reduces to Eq. (S24) for

$$\begin{aligned} k &= 1 \\ a_x &= a_y^{-1} = a_{old} \end{aligned}, \quad \text{(S26)}$$

as we wanted to demonstrate.

### (E) Derivation of an expression for the point $\tau$ separating the low-number and high-number decision behaviours, and proof of the approximate $\Delta N$ rule for low $N$.

We now consider the general expression of the probability, Eq. S22,

$$P_x = \left(1 + \frac{1 + a_x s^{-(n_x - k n_y)}}{1 + a_y s^{-(n_y - k n_x)}}\right)^{-1}.$$
(S27)

For the reasons described below, the transition between the two regimes takes place when the following conditions are met

$$a_x s^{-(n_x - k n_y)} = 1$$

$$a_y s^{-(n_y - k n_x)} = 1.$$
(S28)

These conditions define a point $(\tau_x, \tau_y)$ with

$$\tau_x = \frac{\log(a_x) + k \log(a_y)}{(1 - k^2) \log(s)}$$

$$\tau_y = \frac{k \log(a_x) + \log(a_y)}{(1 - k^2) \log(s)},$$
(S29)

see **Figure S13.**

This transition point is relevant because when the left-hand-side terms of Equation (S28) are much lower than 1 they can be neglected, so $P_x$ is always 0.5. Therefore, the region above the transition point $(\tau_x, \tau_y)$ in which both left-hand-side terms of Equation (S28) are lower than 1 (region 1 in **Figure S14**) is the plateau of $P_x = 0.5$.

On the other hand, if the two left-hand-side terms of Equation (S28) are much higher than 1, we can use the approximations

$1 + a_x s^{-(n_x - k n_y)} \approx a_x s^{-(n_x - k n_y)}$ and $1 + a_y s^{-(n_y - k n_x)} \approx a_y s^{-(n_y - k n_x)}$, to write Eq. (S27) as

$$P_x \approx \left(1 + a_x / a_y s^{-\Delta N(1+k)}\right)^{-1},$$
(S30)

that only depends on $\Delta N$.. Therefore, the region below the transition point $(\tau_x, \tau_y)$ in which both left-hand-side terms of Equation (S28) are higher than 1 (region 2 in **Figure S14**) corresponds to a $\Delta N$ rule for decision making.

For the case of symmetric non-social information, in which $a \equiv a_x = a_y$, Eq. (S29) reduces to

$$\tau \equiv \tau_x = \tau_y = \frac{\log(a)}{(1-k)\log(s)}.$$ (S31)

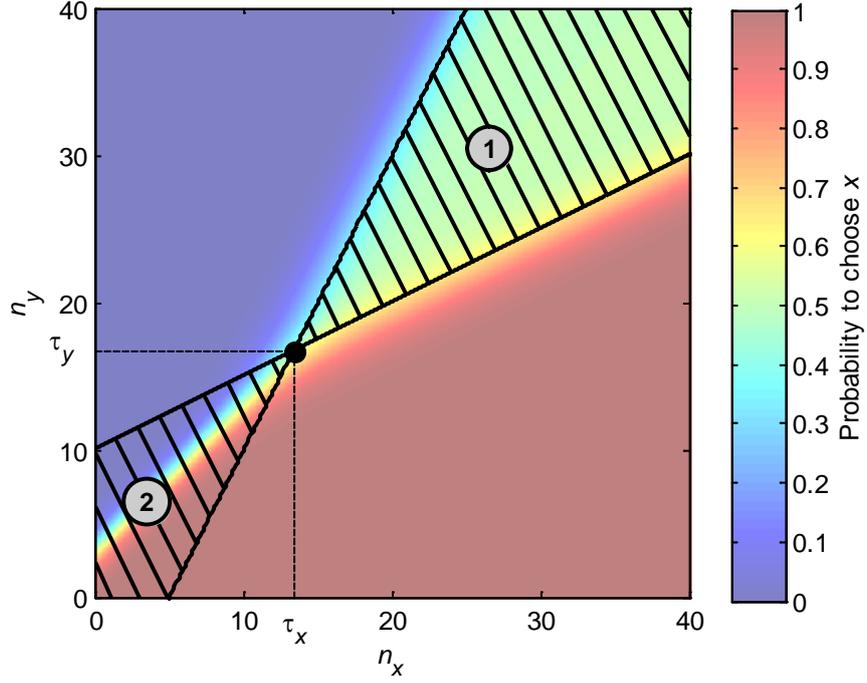

**Figure S14. Transition point $\tau$ between the low and high numbers regimes.** Region 1 corresponds to the plateau with $P_x=0.5$. The $\Delta N$ rule is approximately valid in region 2. Parameters are as in **Fig. S13**.

23